\documentclass[12pt]{article}
\usepackage{putex}
\usepackage{graphicx}

\begin{document}

\preprint{PUPT-2385}

\institution{ENS}{${}^1$Ecole Normale Sup\'erieure, 45 rue d'Ulm, 75005, Paris, France}
\institution{PU}{${}^2$Joseph Henry Laboratories, Princeton University, Princeton, NJ 08544, USA}
\institution{TN}{${}^3$Department of Physics, Technion, Haifa 32000, Israel}

\title{Absence of a Fermi surface in classical minimal \\ four-dimensional gauged supergravity}

\authors{Raphael Belliard,${}^\ENS$ Steven S.~Gubser,${}^\PU$ and Amos Yarom${}^{\PU,\TN}$}

\abstract{
We demonstrate that the two point function of the supercurrent dual to the gravitino in the four-dimensional extremal anti-de Sitter Reissner-Nordstrom black hole does not exhibit a Fermi surface singularity. In our analysis, we utilize the ingoing Eddington-Finkelstein coordinate system, which enables us to bypass certain complications in the determination of the allowed near horizon behavior of the gravitino field at zero frequency.  We check that our method agrees with previous results for the massless charged Dirac field.}

\date{June 2011}

\maketitle

\section{Introduction}
\label{INTRODUCTION}

Considerable interest has attached to the discovery \cite{Liu:2009dm,Cubrovic:2009ye} that two-point functions of spinorial operators, calculated using the gauge-gravity duality applied to an extremal charged black hole background, exhibit singularities at zero frequency and non-zero momentum.  These singularities have been argued to indicate a Fermi surface in the conformal field theory dual to the black hole.  The conformal field theory is subjected to non-zero chemical potential for a global $U(1)$ symmetry, but it is at zero temperature and otherwise undeformed.  The global $U(1)$ becomes a $U(1)$ gauge field in the bulk, and the two-point function of interest is based on the propagation of a charged Dirac fermion in the bulk geometry.

An obvious question, as yet unanswered, is how to derive a bulk action from string theory that admits an appropriate charged black hole background and has a fermion field whose dual two-point function exhibits a Fermi surface singularity. The closest approach to such a top-down construction has been \cite{Ammon:2010pg}, which treats fermions based on probe branes which exhibit $p$-wave superfluidity.  Here we aim for a conceptually simpler construction: We want to find out whether a Fermi surface exists in minimal four-dimensional gauged supergravity.  This theory can certainly be embedded in string theory; for a recent discussion, see \cite{Bah:2010cu}.  The only fermion field in minimal four-dimensional gauged supergravity is a complex gravitino, and it is charged under a $U(1)$ gauge field that, along with the gravitino, completes the graviton supermultiplet.

The answer we find is that there is no Fermi surface singularity in the two-point function of the operator dual to the gravitino.  This operator is the supercurrent of the dual superconformal field theory.  In order to streamline our computations, we consider only the zero-temperature black hole geometry.  Moreover, for the most part we restrict our attention to zero frequency.  A singularity in the supercurrent Green's function, if it exists, should manifest itself as a normal mode of the gravitino.  For a similar computation in a charged $AdS_5$ black hole background, see \cite{Gubser:2009qt}.

In order to obtain the retarded Green's function at non-zero frequency, one must impose infalling boundary conditions at the horizon for the appropriate bulk fields.  For $\omega=0$, it is more subtle to determine the correct horizon boundary conditions.
The approach we favor is to use ingoing Eddington-Finkelstein coordinates for the black hole background.  These coordinates are non-singular at the horizon.  Thus we can be confident that the appropriate boundary conditions at the horizon are to require that the fermion field is itself non-singular at the horizon.

The organization of the rest of this paper is as follows.  In section~\ref{BACKGROUND} we present the bosonic background and summarize our conventions.  In section~\ref{DIRAC} we demonstrate our method by finding the normal mode of a massless charged Dirac field in the extremal anti-de Sitter Reissner-Nordstrom black hole background (hereafter AdSRN).  This normal mode occurs at the same value of $k_F$ found in \cite{Liu:2009dm}.  In section~\ref{GRAVITINO} we explain the gravitino calculation.  We end with a discussion in section~\ref{DISCUSSION}.

While this paper was in progress, we received \cite{Gauntlett:2011mf}, which goes considerably further than we have done in analyzing the two-point correlator of the supercurrent, and also finds no Fermi surface singularities.

\section{Bosonic background}
\label{BACKGROUND}

Throughout, we will work with the following bosonic lagrangian in four dimensions:
 \eqn{BosonicL}{
  {\cal L}_b = {1 \over 4} R - {1 \over 4} F_{\mu\nu}^2 + {3 \over 2L^2} \,.
 }
Our conventions are essentially those of \cite{Romans:1991nq}; in particular, the metric signature is mostly plus.  The AdSRN black hole is the following solution of the classical equations of motion following from \eno{BosonicL}:
 \eqn{AdSRN}{
  ds^2 = -U(r)^2 dv^2 + 2 dv dr + W(r) (dx_1^2 + dx_2^2) \qquad\qquad
  A_\mu dx^\mu = \Phi(r) dv \,,
 }
where $x^\mu = (v,r,x^1,x^2)$ and
 \eqn{UWPhi}{
  U(r) &= \sqrt{{r^2 \over L^2} - {2M \over r} + {Q^2 \over r^2}}  \cr
  W(r) &= {r \over L}  \cr
  \Phi(r) &= Q \left( {1 \over r} - {1 \over r_H} \right) \,,
 }
and $r=r_H$ is the most positive root of the equation 
\begin{equation}
\label{E:Urzero}
	U(r)^2 = 0\,.
\end{equation}
We will require that the black hole is extremal. Thus, $r=r_H$ is a double root of \eqref{E:Urzero}.  One can show that in the extremal limit
 \eqn{MQvalues}{
  M = {2 r_H^3 \over L^2} \qquad\qquad Q = {\sqrt{3} r_H^2 \over L} \,.
 }
We will further simplify our calculations by setting $r_H=L=1$.  This does not cause any essential loss of generality: $L$ is related to the number of degrees of freedom in the dual superconformal field theory, and its value does not affect the location of the Fermi surface. Also, $r_H$ can be altered by rescaling the radial coordinate.  In field theory terms, there is essentially only one state under discussion, namely the superconformal field theory at zero temperature and non-zero chemical potential.  The value of the chemical potential sets a scale against which all other dimensionful quantities, in particular the Fermi momentum, can be measured.

\section{Dirac fermion}
\label{DIRAC}

Before discussing fermions, we must first introduce conventions for spin structure.  We define a vierbein $e_\mu^m$ and a flat metric $\eta_{mn} = \diag\{ -1,1,1,1 \}$ such that $g_{\mu\nu} = \eta_{mn} e^m_\mu e^n_\nu$.  The particular vierbein we use for the metric \eno{AdSRN} is
 \eqn{Vierbein}{
  e^v_\mu dx^\mu = -U(r) dv + {dr \over U(r)} \qquad
  e^r_\mu dx^\mu = {dv \over U(r)} \qquad
  e^{x^1}_\mu dx^\mu = W(r) dx^1 \qquad
  e^{x^2}_\mu dx^\mu = W(r) dx^2 \,,
 }
where the upper indices are tangent space indices.  We will hereafter favor the use of tangent space indices $m$, $n$ over curved space indices $\mu$, $\nu$.  For example, when taking partial derivatives we will use $\partial_m \equiv e^\mu_m \partial_\mu$ instead of $\partial_\mu$.  The standard spin connection $\omega_{mpq}$ is antisymmetric in $p$ and $q$.

We use the following basis of gamma matrices, defined with upper tangent space indices:
\begin{align}
\begin{split}
\label{E:defgamma}
	\gamma^t &= 
	\begin{pmatrix} 
		i & 0 & 0 & 0 \\
		0 & i & 0 & 0 \\
		0 & 0 & -i & 0 \\
		0 & 0 & 0 & -i
	\end{pmatrix}
	\qquad
	\gamma^r=
	\begin{pmatrix}
		0 & 0 & 0 & -i \\
		0 & 0 & -i & 0 \\
		0 & i & 0 & 0 \\
		i & 0 & 0 & 0 
	\end{pmatrix}
	\\
	\gamma^{x^1} &= 
	\begin{pmatrix} 
		0 & 0 & 0 &-1 \\
		0 & 0 & 1 & 0 \\
		0 & 1 & 0 & 0 \\
		-1 & 0 & 0 & 0 
	\end{pmatrix}
	\qquad
	\gamma^{x^2}=
	\begin{pmatrix}
		0 & 0 & -i & 0 \\
		0 & 0 & 0 & i \\
		i & 0 & 0 & 0 \\
		0 & -i & 0 & 0 
	\end{pmatrix} \,.
\end{split}
\end{align}
We define anti-symmetrization of indices so that $[mn] = {1 \over 2} (mn-nm)$ and $[mnp] = {1 \over 6} (mnp+pmn+npm-nmp-pnm-mpn)$, and we follow standard the standard notation $\gamma^{mn} = \gamma^{[m} \gamma^{n]}$ and $\gamma^{mnp} = \gamma^{[m} \gamma^n \gamma^{p]}$.

To demonstrate the usefulness of the ingoing Eddington-Finkelstein coordinate system \eqref{AdSRN}  in a simple context familiar from the literature, let us augment the bosonic action \eno{BosonicL} by the action for a massless, charged Dirac fermion:
 \eqn{LDirac}{
  {\cal L}_{D} = -\overline\psi \gamma^m D_m \psi
 }
where $\overline\psi = i\psi^\dagger \gamma^t$ and
 \eqn{DDef}{
  D_m \psi = \partial_m \psi + {1 \over 4} \omega_{mpq} \gamma^{pq} \psi - i q A_m \psi \,.
 }
To solve the equations of motion $\gamma^m D_m \psi = 0$, we first make the ansatz
 \eqn{DiracFourier}{
  \psi = e^{-i\omega v + i k x^1} \psi(r) \,.
 }
Next we observe that the equations of motion partially decouple, leading to
 \eqn{GroupADirac}{
  \psi_1' + \left( {U' \over 2U} + {W' \over W} \right) \psi_1 &= 
    {k \over UW} \psi_1 + {i (\omega + q\Phi) \over U^2} (\psi_1 + \psi_4)  \cr
  \psi_4' + \left( {U' \over 2U} + {W' \over W} \right) \psi_4 &= 
    -{k \over UW} \psi_4 + {i (\omega + q\Phi) \over U^2} (\psi_1 + \psi_4) \,,
 }
and a similar pair of equations for $\psi_2$ and $\psi_3$, obtained from \eno{GroupADirac} through the replacements $\psi_1 \to \psi_2$, $\psi_4 \to \psi_3$, and $k \to -k$.  For the sake of brevity, from here on we will consider only solutions where $\psi_2=\psi_3=0$.

Because the Dirac field is massless, in an asymptotically $AdS_4$ geometry all solutions to \eno{GroupADirac} behave close to the boundary as $\psi_\alpha \propto r^{-3/2}$.  It is therefore delicate to define which solutions are ``normalizable,'' i.e., which solutions are associated with the expectation value of a spinorial operator in the dual field theory.  The resolution (well understood in previous literature, for example \cite{Liu:2009dm}) is that solutions of the classical equations of motion which are eigenvectors of $\gamma^r$ with eigenvalue $+1$ are considered to be sources for the dual spinorial operator, while eigenvectors with eigenvalue $-1$ are considered to describe expectation values of this operator.  The latter solutions should be regarded as normalizable.  With our choice of gamma matrices, the normalizable solutions are the ones where $\lim_{r \to \infty} \psi_1(r)/\psi_4(r) = i$.

Near the horizon, where the geometry is $AdS_2 \times {\bf R}^2$, one may straightforwardly show that the two solutions of \eno{GroupADirac} take the form
 \eqn{psiNear}{
  \psi = u_\pm (r-1)^{\nu_\pm} \qquad\hbox{where}\qquad
   \nu_\pm = -{1 \over 2} - {iq \over 2\sqrt{3}} \pm {1 \over 6} \sqrt{6k^2 - 3q^2} \,,
 }
where we have set $r_H=L=1$.  The form of the eigenspinors $u_\pm$ is not enlightening and we will not write them out explicitly.  The point of our analysis in Eddington-Finkelstein coordinates is that one can immediately rule out the $\nu_-$ solution on grounds that it diverges at the future horizon: recall that the Eddington-Finkelstein coordinates are themselves non-singular at the horizon, so a divergence in $\psi$ is a signal of genuinely singular behavior.  Thus we must use the $\nu_+$ solution in constructing a normal mode.  For any given $k$ and $q$, it is straightforward to integrate the equations \eno{GroupADirac} numerically, starting with the $\nu_+$ solution very close to $r=1$, and proceed out to a large value of $r$, where $\psi_1(r)/\psi_4(r)$ may be evaluated and compared to the desired value of $i$ for a normalizable mode.  In order to compare with \cite{Liu:2009dm} we chose $q=1$, and we found that a normal mode arises at $k_F = 0.918$, consistent with the Fermi momentum found in that work.

\section{The gravitino}
\label{GRAVITINO}

The gravitino action for minimally gauged supergravity was constructed in \cite{Freedman:1976aw} and is given by
\begin{equation}
\label{E:gaction}
	\mathcal{L}_g = -\frac{1}{2} \overline{\psi}_m \gamma^{mnp} D_n \psi_p  + \frac{1}{2L} \overline{\psi}_m \gamma^{mn} \psi_{n}  - 
	\frac{i}{4} \overline\psi_m F^{pq} \gamma_p \gamma^{mn} \gamma_q \psi_n + 
	\left( {\hbox{\small $4$-fermi interactions}}\right)\,.
\end{equation}
The Lagrangian \eqref{E:gaction} together with \eqref{BosonicL} is the minimal $\mathcal{N}=2$ gauged supergravity Lagrangian in four dimensions.
In writing \eqref{E:gaction} we have closely followed the notation of \cite{Romans:1991nq}: in particular, $\psi_{m\,\alpha}$ is a complex spin-$3/2$ field, and $\alpha=1,\ldots,4$ are spinor indices which we have suppressed when writing the action \eqref{E:gaction}.  Conventions for spinors and tangent space indices are as established in section~\ref{DIRAC}.
The covariant derivative is defined through
\begin{equation}
	D_{m} \psi_{n} = \partial_{m} \psi_{n} - \omega_{m\phantom{p}n}^{\phantom{m}p} \psi_{p} + \frac{1}{4} \omega_{m pq}\gamma^{pq} \psi_n - i A_{m} \psi_{n}\,.
\end{equation}
As before, we set $L=r_H=1$.

For the purpose of computing the two point function of the supercurrent it is sufficient to consider the equations of motion to linear order in $\psi$,
\begin{equation}
\label{E:RSEOM}
	\gamma^{mnp} D_n \psi_p - \gamma^{mn} \psi_{n} + \frac{i}{2} F^{pq} \gamma_p \gamma^{mn} \gamma_q \psi_n = 0.
\end{equation}
After using our conventions \eqref{E:defgamma} for the gamma matrices, and using an ansatz 
\begin{equation}
	\psi_m = e^{-i\omega t + i k x^1}\psi_m(r)
\end{equation}	
the gravitino equations of motion take the schematic form
\begin{equation}
\label{E:RSEOMMatrix}
	\mathbf{K} \frac{d}{d r} \begin{pmatrix}
		\psi_t \\
		\psi_x \\
		\psi_y
	\end{pmatrix}
	+
	\mathbf{M}'
	\begin{pmatrix}
		\psi_t \\
		\psi_x \\
		\psi_y \\
		\psi_r
	\end{pmatrix} 
	=0\,,
\qquad
	\mathbf{M}_c' 
	\begin{pmatrix}
		\psi_t \\
		\psi_x \\
		\psi_y 
	\end{pmatrix} 
	=0\,.
\end{equation}
Here $\mathbf{K}$ is a $12 \times 12$ matrix, $\mathbf{M}'$ is a $12 \times 16$ matrix and $\mathbf{M}_c'$ is a $4 \times 12$ matrix. We have used a prime on the matrices $\mathbf{M}'$ and $\mathbf{M}_c'$ to distinguish them from the matrices $\mathbf{M}$ and $\mathbf{M}_c$ which we will introduce shortly. Overall, we have sixteen equations: twelve differential equations and four algebraic equations. The matrix $\mathbf{K}$ is invertible and, due to its simple form, it is straightforward to obtain an explicit expression for its inverse. 

We have not yet fixed the superdiffeomorphism invariance of \eqref{E:RSEOMMatrix}. Indeed, 
\begin{equation}
	\psi_m = \partial_m \epsilon + \frac{1}{4}\omega_{mpq}\gamma^{pq} - i A_m \epsilon + \frac{1}{2} \gamma_m \epsilon + i \frac{1}{4} F_{mn}\gamma^{mn} \epsilon
\end{equation}
is a solution to the equations of motion which is gauge equivalent to $\psi_m=0$. To partially fix this gauge we set
\begin{equation}
\label{E:gtgauge}
	\gamma_m \psi^m = 0
\end{equation}
in order to fix the four components of $\psi_r$ in terms of the twelve components of $\psi_t$, $\psi_{x^1}$ and $\psi_{x^2}$. We found this gauge choice more convenient than an axial gauge $\psi_r=0$ which is, perhaps, more obviously in the spirit of the AdS/CFT duality.  After implementing the gauge choice \eqref{E:gtgauge} the equations of motion can be written in the block diagonal form:
\begin{equation}
\label{E:EOM}
	\frac{d}{d r} \begin{pmatrix} 
		\Psi \\ \widetilde{\Psi}
		\end{pmatrix}
	=
	\begin{pmatrix}
		\mathbf{M} & 0 \\
		0 & \mathbf{\widetilde{M}} 
	\end{pmatrix}
	\begin{pmatrix} 
		\Psi \\ \widetilde{\Psi}
	\end{pmatrix}	
	\qquad
	\begin{pmatrix}
		\mathbf{M}_c & 0 \\
	0 & \mathbf{\widetilde{M}}_c 
	\end{pmatrix}
	\begin{pmatrix} 
		\Psi \\ \widetilde{\Psi}
	\end{pmatrix}
	 = 0
\end{equation}
with 
\begin{equation}
	\Psi = \begin{pmatrix}
		\psi_{t1} &
		\psi_{t4} & 
		\psi_{x1} &
		\psi_{x4} &
		\psi_{y2} &
		\psi_{y3} 
	\end{pmatrix}
	\qquad
	\widetilde{\Psi} = \begin{pmatrix}
		\psi_{t2} &
		\psi_{t3} &
		\psi_{x2} &
		\psi_{x3} &
		\psi_{y1} &
		\psi_{y4} 
	\end{pmatrix}
\end{equation}
and $\mathbf{M}$ and $\mathbf{M}_c$ are given by
\begin{equation}
U \mathbf{M} = 
	\begin{pmatrix}
		M_{+}(k) &
		\frac{i}{2}-\frac{i W_+}{U} &
		-\frac{W_-}{2 U} &
		\frac{i U_+}{4 r} &
		-\frac{i W_-}{2 U} &
		-\frac{2 k+U_+}{4 r} \\
		\frac{3 i}{2}-\frac{i W_+}{U} &
		M_-(k) &
		-\frac{i U_-}{4 r} &
		\frac{W_+}{2 U} &
		\frac{U_--2 k}{4 r} &
		\frac{i W_+}{2 U} \\
		\frac{1}{2} &
		-\frac{i K_-}{2 r}&
		M_0(4k)&
		-\frac{i W_+}{2 U}&
		-\frac{i \left(2 k-2 (r U)' +U_-\right)}{4 r}&
		-\frac{W_+}{2 U} \\
		-\frac{i K_+}{2 r}&
		\frac{1}{2}&
		-\frac{i W_-}{2 U}&
		M_0(-4k)&
		-\frac{W_-}{2 U}&
		\frac{i \left(2 k+2 (r U)' -U_+\right)}{4 r} \\
		-\frac{i}{2}&
		-\frac{K_-}{2 r}&
		\frac{i \left(U_- -2 (r U)' \right)}{4 r}&
		\frac{W_+}{2 U}&
		M_0(-2k)&
		-\frac{i W_+}{2 U}\\
		 -\frac{K_+}{2 r}&
		 -\frac{i}{2}&
		 \frac{W_-}{2 U}&
		 \frac{i \left(U_+ -2 (r U)' \right)}{4 r}&
		 -\frac{i W_-}{2 U}&
		 M_0(2k)\\
   	\end{pmatrix}
\end{equation}
and
\begin{equation}
M_c = 
	\begin{pmatrix}
		-\frac{i K_- U}{r} &
		U&
		i W_-&
		-\frac{U \left(r U\right)'}{2 r}&
		-W_-&
		-\frac{i U \left(-2 k+(r U)'\right)}{2 r} \\
		U&
		-\frac{i K_+ U}{r}&
		-\frac{U \left(r U\right)'}{2 r}&
		i W_+&
		-\frac{i U \left(2 k+(r U)'\right)}{2 r}&
		-W_+
	\end{pmatrix}
\end{equation}
with
\begin{align}
\begin{split}
	M_{\pm}(k) &= \mp \frac{k}{2r} - \frac{i\omega}{U} + \frac{1}{2} \left(\frac{U}{r} - \frac{2 i \Phi}{U} + 3 U' \pm  \Phi'\right)
	\qquad
	M_0(k) = \frac{7 U^2 - 4 i r (\omega+\Phi) + U (r U' - k)}{4 r U} 
	\\
	U_{\pm} & = - U + 3 r U' \pm 4 r \Phi'
	\qquad
	W_{\pm} = \omega \pm U + \Phi 
	\qquad
	K_{\pm} = k \pm U -  r \Phi'\,.
\end{split}
\end{align}
The kinematic equations of motion for $\widetilde{\Psi}$ can be obtained from those of $\Psi$ by making the replacement
$\psi_{t1} \to \psi_{t2}$, $\psi_{t4} \to \psi_{t3}$, $\psi_{x1}\to -\psi_{x2}$, $\psi_{x4} \to - \psi_{x3}$, $\psi_{y2} \to -\psi_{y1}$, $\psi_{y3} \to -\psi_{y4}$ and taking $k \to  -k$. The constraint equations for $\widetilde{\Psi}$ can be obtained from those for $\Psi$ by making the replacement $\psi_{t1} \to -\psi_{t3}$, $\psi_{t4} \to \psi_{t2}$, $\psi_{x1}\to \psi_{x3}$, $\psi_{x4} \to - \psi_{x2}$, $\psi_{y2} \to \psi_{y4}$, $\psi_{y3} \to -\psi_{y1}$ and taking $\omega \to  -\omega$ and $\Phi \to -\Phi$.
	
We focus on the equations of motion for $\Psi$. The equations of motion for $\widetilde{\Psi}$ can be treated in an equivalent manner. Near the asymptotically AdS boundary located at $r \to \infty$ we find that \eqref{E:EOM} admits two linearly independent solutions,
\begin{equation}
\label{E:Sources}
	\Psi_{b\,1} = r^{-\frac{1}{2}}  \begin{pmatrix} -i \\ -1 \\ i \\ 1 \\ 0 \\ 0 \end{pmatrix}  + \mathcal{O}(r^{-1}) 
	\qquad
	\Psi_{b\,2} =  r^{-\frac{1}{2}} \begin{pmatrix} -i \\ -1 \\ 0 \\ 0 \\ i \\ 1 \end{pmatrix}  + \mathcal{O}(r^{-1})
\end{equation}	
so that the most general near boundary asymptotics take the form
\begin{equation}
	\Psi = S_1 \Psi_{b\,1} + S_2 \Psi_{b\,2}\,.
\end{equation}
In addition to \eqref{E:Sources} there are two more solutions which control the coefficients of the $r^{-5/2}$ terms in a series expansion for $\Psi$. The coefficients $S_1$ and $S_2$ correspond to the source terms for the supercurrent while the coefficients of the $r^{-5/2}$ terms in a near boundary expansion for $\Psi$ control the expectation value of the supercurrent (see, for example, \cite{Volovich:1998tj,Policastro:2008cx}). 

Near the horizon of the extremal black hole described in section \ref{BACKGROUND}, the spin-$3/2$ field admits four linearly independent solutions
\begin{equation}
\label{E:NHExpansion}
	\Psi_{\pm\,a} = (r-1)^{\nu_{\pm}} (1+O(r-1)) u_{\pm\,a} \qquad \hbox{where} \qquad \nu_{\pm} = -\frac{1}{2} + \frac{i}{2 \sqrt{3}} \pm \frac{1}{6} \sqrt{21+6 k^2}
\end{equation}
and $a=1,2$. We omit the rather long expressions for $u_{\pm\,1}$ and $u_{\pm\,2}$.
We are working in a coordinate patch which is continuous across the event horizon of the black hole. Therefore, we do not expect physical solutions to diverge at $r=1$ and so, we must discard the solutions $\Psi_-$. Thus, the most general solution to the equations of motion take the form:
\begin{equation}
	\Psi = H_1 \Psi_{+\,1} + H_2 \Psi_{+\,2}
\end{equation}
with $H_1$ and $H_2$ integration constants.

\begin{figure}
\begin{center}
\includegraphics{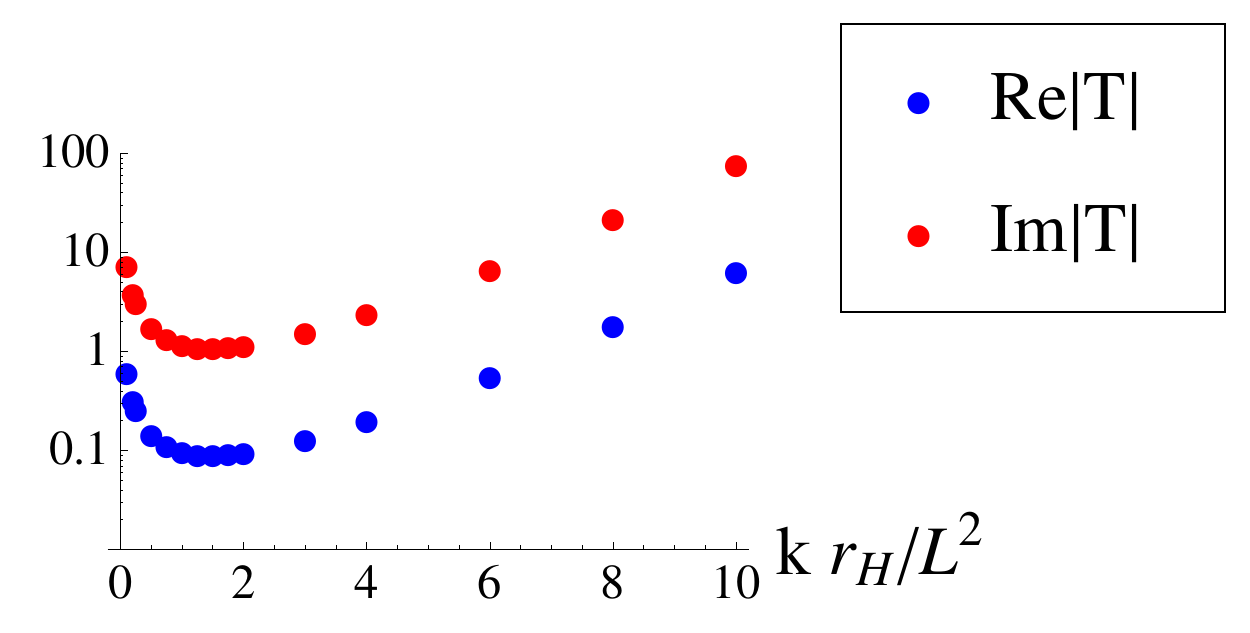}
\caption{\label{F:detT} (Color online) A plot of the numerically evaluated real and imaginary parts of the determinant of the transformation matrix $T$ defined in \eqref{E:Tdef}. A Fermi surface exists when the determinant vanishes. We have checked that $|T|$ does not vanish up to $k r_H/L^2 = 10$.}
\end{center}
\end{figure}

Recall that we are interested in poles of the retarded supercurrent-supercurrent correlator at finite momentum $k$ and zero frequency. A pole of the correlator implies a possible excitation of the supercurrent with no source being applied. From the point of view of the gravitino dual such a configuration would involve a non-trivial solution for which $S_1 = S_2 = 0$. More precisely, the integration constants $S_1$ and $S_2$ are linearly related to the near horizon behavior of the solution  through
\begin{equation}
\label{E:Tdef}
	\begin{pmatrix}
		S_1 \\ S_2 
	\end{pmatrix}
	=
	T
	\begin{pmatrix}
		H_1 \\ H_2 \,
	\end{pmatrix}\,.
\end{equation}	
The condition that there exist a non trivial solution for which $S_1=S_2=0$ amounts to the condition that the determinant of $T$ vanishes:
\begin{equation}
	| T |= 0\,.
\end{equation}
It is straightforward to integrate \eqref{E:EOM} numerically to obtain the matrix $T$ defined in \eqref{E:Tdef}. In figure \ref{F:detT} we have plotted $| T |$ as a function of $k$. We have checked that a Fermi surface does not exist all the way to $k r_H/L^2 = 10$ including both the $\Psi$ and the $\widetilde{\Psi}$ components of $\psi_m$. 

\section{Discussion}
\label{DISCUSSION}

The absence of a Fermi surface singularity in the two-point function of the supercurrent is a striking contrast to the well-known result that such singularities do appear for operators dual to Dirac fermions of appropriate charge.  There are two possible views one could take on the significance of these results:
 \begin{enumerate}
  \item On one hand, one could reason that all the results showing the existence of a Fermi surface, with the notable exception of the probe brane calculations of \cite{Ammon:2010pg}, have been based on {\it ad hoc} fermionic actions.  So---one might argue---there isn't yet a compelling reason to believe that real string theory constructions based only on bulk dynamics (not probe branes) exhibit Fermi surface behavior.  In further support of this view, one could note that since the supercurrent is a superpartner of the stress tensor, it couples to most of the fields in the the dual field theory; thus if there is a feature like a Fermi surface, it should be visible in correlators of the supercurrent.  Moreover, although minimal four-dimensional gauged supergravity is not a complete theory in itself, it can be embedded in many of the string theory constructions which admit an AdSRN black hole.  Altogether, the most aggressive claim that could be made is that absence of gravitino normal modes in the extremal AdSRN black hole background constitutes preliminary evidence that there is no Fermi surface in the dual field theory.
  \item On the other hand, one could hypothesize, along the lines of \cite{Huijse:2011hp}, that some fraction of the fundamental fermions in the dual field theory become bound with fundamental bosons, and the resulting fermionic bound states exhibit a Fermi surface; but there are some unpaired fermions left over which experience strong gauge interactions and as a result do not exhibit a Fermi surface.  On this interpretation, what one is seeing in the two-point function of the supercurrent is the absence of a Fermi surface for the unpaired fermions.  This makes sense because there are effectively ${\cal O}(N^{3/2})$ species of unpaired fermions, where $N$ is the number of M2-branes in an underlying M-theory construction, and a correlator computed from classical gravitino propagation captures the leading order ${\cal O}(N^{3/2})$ behavior of the Green's functions.  There are only ${\cal O}(1)$ species of paired fermions, so to see their Fermi surface effects, one would need to investigate loop corrections to gravitino propagation.  It may be supposed then that the two-point function computed from some appropriate Dirac fermion would directly exhibit the Fermi surface behavior of the paired fermions.  In short, the most optimistic claim that could be made is that no Fermi surface was expected for the gravitinos at the classical level, making our results a confirmation of the overall picture of \cite{Huijse:2011hp}.
 \end{enumerate}
In balancing these competing interpretations, we prefer a degree of agnosticism.  Clearly there is no Fermi surface visible from classical computations in the AdSRN black hole background of minimal four-dimensional gauged supergravity.  The AdSRN black hole background presents other puzzles, including zero-point entropy and stability issues.  Absent a clearer field theoretic account that resolves the zero-point entropy issue, the best approach to deciding between the two interpretations presented would be to find out whether other top-down calculations do exhibit Fermi surface singularities, and to have control over prefactors that reveal the $N$-dependence of the number of species participating in the singularity.

\section*{Acknowledgments}

We are grateful to Matthias Kaminski for helpful discussions.  This work was supported in part by the Department of Energy under Grant No.~DE-FG02-91ER40671.

\bibliographystyle{JHEP}
\bibliography{gravitino}

\providecommand{\href}[2]{#2}\begingroup\raggedright\begin{thebibliography}{10}

\bibitem{Liu:2009dm}
H.~Liu, J.~McGreevy, and D.~Vegh, ``{Non-Fermi liquids from holography},''
  {\em Phys. Rev.} {\bf D83} (2011) 065029,
  [\href{http://xxx.lanl.gov/abs/0903.2477}{{\tt arXiv:0903.2477}}].

\bibitem{Cubrovic:2009ye}
M.~Cubrovic, J.~Zaanen, and K.~Schalm, ``{String Theory, Quantum Phase
  Transitions and the Emergent Fermi-Liquid},''  {\em Science} {\bf 325} (2009)
  439--444, [\href{http://xxx.lanl.gov/abs/0904.1993}{{\tt arXiv:0904.1993}}].

\bibitem{Ammon:2010pg}
M.~Ammon, J.~Erdmenger, M.~Kaminski, and A.~O'Bannon, ``{Fermionic Operator
  Mixing in Holographic p-wave Superfluids},''  {\em JHEP} {\bf 05} (2010) 053,
  [\href{http://xxx.lanl.gov/abs/1003.1134}{{\tt arXiv:1003.1134}}].

\bibitem{Bah:2010cu}
I.~Bah, A.~Faraggi, J.~I. Jottar, and R.~G. Leigh, ``{Fermions and Type IIB
  Supergravity On Squashed Sasaki- Einstein Manifolds},''  {\em JHEP} {\bf 01}
  (2011) 100, [\href{http://xxx.lanl.gov/abs/1009.1615}{{\tt
  arXiv:1009.1615}}].

\bibitem{Gubser:2009qt}
S.~S. Gubser and F.~D. Rocha, ``{Peculiar properties of a charged dilatonic
  black hole in $AdS_5$},''  {\em Phys. Rev.} {\bf D81} (2010) 046001,
  [\href{http://xxx.lanl.gov/abs/0911.2898}{{\tt arXiv:0911.2898}}].

\bibitem{Gauntlett:2011mf}
J.~P. Gauntlett, J.~Sonner, and D.~Waldram, ``{The spectral function of the
  supersymmetry current (I)},''  \href{http://xxx.lanl.gov/abs/1106.4694}{{\tt
  arXiv:1106.4694}}.

\bibitem{Romans:1991nq}
L.~Romans, ``{Supersymmetric, cold and lukewarm black holes in cosmological
  Einstein-Maxwell theory},''  {\em Nucl.Phys.} {\bf B383} (1992) 395--415,
  [\href{http://xxx.lanl.gov/abs/hep-th/9203018}{{\tt hep-th/9203018}}].

\bibitem{Freedman:1976aw}
D.~Z. Freedman and A.~K. Das, ``{Gauge Internal Symmetry in Extended
  Supergravity},''  {\em Nucl.Phys.} {\bf B120} (1977) 221.

\bibitem{Volovich:1998tj}
A.~Volovich, ``{Rarita-Schwinger field in the AdS / CFT correspondence},''
  {\em JHEP} {\bf 9809} (1998) 022,
  [\href{http://xxx.lanl.gov/abs/hep-th/9809009}{{\tt hep-th/9809009}}].

\bibitem{Policastro:2008cx}
G.~Policastro, ``{Supersymmetric hydrodynamics from the AdS/CFT
  correspondence},''  {\em JHEP} {\bf 0902} (2009) 034,
  [\href{http://xxx.lanl.gov/abs/0812.0992}{{\tt arXiv:0812.0992}}].

\bibitem{Huijse:2011hp}
L.~Huijse and S.~Sachdev, ``{Fermi surfaces and gauge-gravity duality},''
  \href{http://xxx.lanl.gov/abs/1104.5022}{{\tt arXiv:1104.5022}}.

\end{thebibliography}\endgroup

\end{document}